\documentclass{elsart}
\usepackage{graphicx}

\begin{document}

\begin{frontmatter}
  \title{Effects of temperature and surface step on the incipient plasticity in strained aluminium studied by atomistic simulations}
  \author[LMP]{P.~Hirel\corauthref{cor}}
  \author[LMP]{S.~Brochard}
  \author[LMP]{L.~Pizzagalli}
  \author[LMP]{P.~Beauchamp}
  \corauth[cor]{Corresponding author.}
  \address[LMP]{Laboratoire de M\'etallurgie Physique, Bat. SP2MI, Bvd M. et P. Curie BP 30179 86 962 Futuroscope Chasseneuil Cedex, FRANCE}
  \date{}

  \maketitle

\begin{abstract}
  Atomistic simulations using an EAM potential are carried out to investigate the first stages of plasticity in aluminum slabs, in particular the effect of both temperature and step geometry on the nucleation of dislocations from surface steps. Temperature is shown to significantly reduce the elastic limit, and to activate the nucleation of dislocation half-loops. Twinning occurs by successive nucleations in adjacent glide planes. The presence of a kinked step is shown to have no influence on the nucleation mechanisms.
\end{abstract}


\end{frontmatter}


The study of mechanical properties takes a new and more critical aspect when applied to nanostructured materials. While plasticity in bulk systems is related to dislocations multiplying from pre-existing defects, such as Franck-Read sources \cite{Hirth1982}, nanostructured materials are too small for such sources to operate, and their plasticity is more likely initiated by dislocations nucleation from surfaces and interfaces \cite{Albrecht1995,Xu2000,Brochard2000a,Godet2004}. In particular, nucleation from grain boundaries is of great interest for the understanding of elementary mechanisms occuring in work hardening of nano-grained materials \cite{Spearot2005,VanSw2002,Yamakov2001,Yamakov2002}. The mechanisms involving the nucleation of dislocations from crack tips are also of great importance to account for brittle to ductile transition in semiconductors \cite{Cleri1997,Zhou1997,Zhu2004}.

In epitaxially-grown thin films, misfit induces a strain and can lead to the formation of dislocations at interfaces \cite{Ernst1997,Wu2001,Trushin2002}. The presence of defects in a surface, such as steps, terraces or hillocks, can also initiate plasticity \cite{Xu2003}. In particular, experimental and theoretical investigations have established that stress concentration near surface steps facilitates the nucleation of dislocations from these sites \cite{Brochard2000,Zimmerman2001}. Dislocations formation in such nanostructures changes their mechanical, electrical, and optical properties, and then may have a dramatic effect on the behaviour of electronic devices \cite{Carrasco2004}. Hence, the understanding of the mechanisms initiating the formation of dislocations in these nanostructures is of high importance.

Since these mechanisms occur at small spatial and temporal scales, which are difficult to reach experimentally, atomistic simulations are well suited for their study. Face-centered cubic metals are first-choice model materials, because of their ductile behaviour at low temperatures, involving a low thermal activation energies. In addition, the development of semi-empirical potentials for metals has made possible the modelling of large systems, and the accurate reproduction of defects energies and dislocation cores structures. Aluminium is used here as a model material.

In this study we investigate the first stages of plasticity in aluminum f.c.c. slabs by molecular dynamics simulations. Evidence of the role of temperature in the elastic limit reduction and in the nucleation of dislocation half-loops from surface steps is obtained. Steps in real crystals are rarely straight, and it has been proposed that a notch or kinked-step would initiate the nucleation of a dislocation half-loop \cite{Pirouz1995,Edirisinghe1997}. This is investigated here by comparing the plastic events obtained from either straight and non-straight steps.

\begin{figure}

  \centering
  \includegraphics[angle=-90, width=0.5\linewidth]{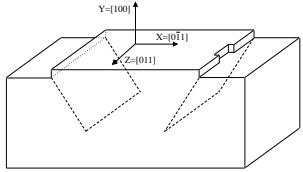}
  \caption{\small{System used in simulations, with periodic boundary conditions along X and Z, and free \{100\} surfaces. The \{111\} glide planes passing through the steps edges are drawn (dashed lines). Here, a notch is built on the right-side step.}}
\label{CrystalGeom}

\end{figure}

Our model consists of a f.c.c. monocrystal, with two \{100\} free surfaces (Fig.~\ref{CrystalGeom}). Periodic boundary conditions are applied along the two other directions, X$=[0 \bar{1} 1]$ and Z$=[011]$. On one \{100\} surface, two opposite, monoatomic steps are built by removing atoms. They lie along Z, which is the intersection between a \{111\} plane and the surface. Such a geometry is therefore well suited to study glide events occuring in \{111\} planes. We investigate tensile stress orthogonal to steps. In this case, Schmid analysis reveals that Shockley partials with a Burgers vector orthogonal to the surface step are predicted to be activated in \{111\} planes in which glide reduces the steps height \cite{Brochard2000a}. In some calculations, consecutive atoms have been removed in the step edge, forming a notch (Fig.~\ref{CrystalGeom}), for investigating the effect of step irregularities on the plasticity. Various crystal dimensions have been considered, from $24\times 16 \times 10$ (3680 atoms), up to $60\times 40 \times 60$ (142800 atoms). The latter crystal size was shown to be large enough to have no influence on the results.

Interactions between aluminum atoms are described by an embedded atom method (EAM) potential, fitted on experimental values of cohesive energy, elastic moduli, vacancy formation and intrinsic stacking fault energies \cite{Pontikis1998}. It is well suited for our investigations since it correctly reproduces the dislocations core structures.

Without temperature, the system energy is minimized using a conjugate-gradient algorithm. The relaxation is stopped when all forces fall below $6.24\times10^{-6}$~eV.\AA$^{-1}$. Then the crystal is elongated by 1\% of its original length along the X direction, i.e. perpendicular to the step. The corresponding strain is applied along the Z direction, according to the isotropic Poisson's ratio of aluminum (0.35). Use of isotropic elasticity theory is justified here by the very low anisotropy coefficient of this material: $A=2C_{44}/(C_{11}-C_{12}) $ = 1.07 (EAM potential used here) ; 1.22 (experiments \cite{Zener1948,Thomas1968}). After deformation, a new energy minimization is performed, and this process is repeated until a plastic event, such as the nucleation of a dislocation, is observed. The occurence of such an event defines the elastic limit of the material at 0K.

At finite temperature, molecular dynamics simulations are performed with the xMD code \cite{XMD2002}, using the same EAM potential. Temperature is introduced by initially assigning an appropriate Maxwell-Boltzmann distribution of atomic velocities, and maintained by smooth rescaling at each dynamics step. The time step is $4\times 10^{-15}$~s, small enough to produce no energy drift during a 300K run. After 5000 steps, ie. 20~ps, the crystal is deformed by 1\%, similarly to what is done at 0K, and then the simulation is continued. If a nucleation event occurs, the simulation is restarted from a previously saved state and using a lower 0.1\% deformation increment.

To visualize formed defects, atoms are colored as a function of a centrosymmetry criterion \cite{Li2003}: atoms not in a perfect f.c.c. environment, ie. atoms on surfaces, in dislocation cores and stacking faults, can then be easily distinguished. In case of dislocation formation, the core position and Burgers vector are determined by computing the relative displacements of atoms in the glide plane. These displacements are then normalized to the edge and screw components of a perfect dislocation.

\begin{figure}

  \centering
  \includegraphics[width=0.5\linewidth]{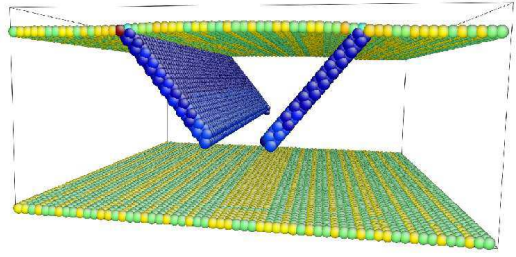}
  \caption{Formation of two dislocations at 0K, after a 10\% elongation of a $60\times 40 \times 60$ crystal. Initial positions of the surface steps are shown (arrows). Only atoms which are not in a perfect f.c.c. environment are drawn: surfaces (yellow-green), stacking fault (dark blue), dislocation cores (light blue) (color online).}
\label{Ech04Atom}

\end{figure}

At 0K, the deformation is found to be purely elastic up to an elongation of 10\%. Then a significant decrease of the total energy suggests an important atomic reorganisation. Crystal vizualisation reveals the presence of defects located in \{111\} planes passing through step edges, and a step height reduction by 2/3 (Fig.~\ref{Ech04Atom}). Atomic displacements analysis in these planes shows that plasticity has occured by the nucleation of dislocations, with Burgers vectors orthogonal to the steps and with a magnitude corresponding to a $90^{\circ}$ partial. This is consistent with the 2/3 reduction of the steps height. The dislocations are straight, the strain being homogeneous all along the steps, and intrinsic stacking faults are left behind. The formation of dislocations from a surface step has already been investigated from 0K simulations, using quasi-bidimensional aluminum crystals \cite{Brochard2000a}. It has been shown that straight $90^{\circ}$ Shockley partials nucleate from the steps. However, the small dimension of the step line did not allow the bending of dislocations. Here, although this restriction has been removed by considering larger crystals (up to 90 atomic planes along Z), only straight partial dislocations have been obtained.

In order to bring the role of steps to light, calculations are performed with a $30\times20\times30$ crystal with two free surfaces, but without step. In that case, plasticity occurs for a much larger elongation, 20\%, and leads to a complex defects structure. It clearly shows the important role played by steps, by significantly reducing the energy barrier due to dislocation-surface interaction, and initiating the nucleation in specific glide planes.

\begin{figure}

  \centering
  \includegraphics[width=\linewidth]{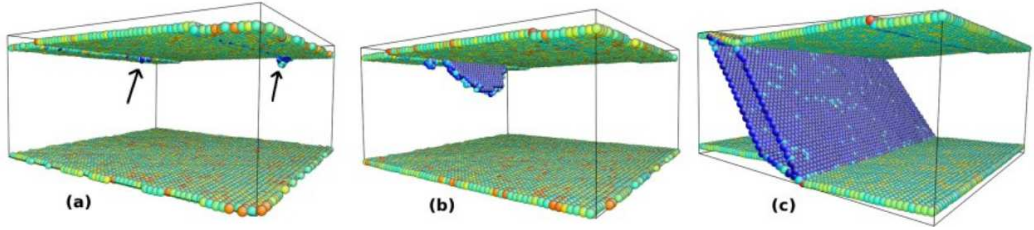}
  \caption{Evolution of the aluminum crystal after a 6.6\% elongation at 300K. Same color convention as Fig.~\ref{Ech04Atom}. The origin of time is when the applied strain is increased to 6.6\%. (a) At 12~ps, several dislocation embryos appeared on both steps (arrows). (b) At 20~ps, a faulted half-loop dislocation has nucleated on one step. (c) After 76~ps, a stable twin was formed. The other step (on the right) remains intact.}
\label{Ech32Atom1}

\end{figure}

\begin{figure}

  \centering
  \includegraphics[angle=-90, width=0.5\linewidth]{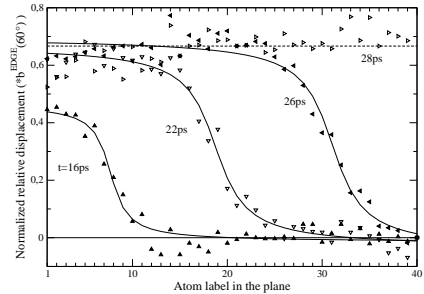}
  \caption{Calculated edge component of the relative displacements of atoms in the activated glide plane and in the Z-layer corresponding to the dislocation front line, at 300K and for different times (triangles). They are fitted with an arctan function (solid lines) according to elasticity theory, for monitoring the dislocation core position during the simulation. The abcissa labels the depth of the atoms from the top surface : 1 corresponds to atoms at the edge of the initial step, and 40 corresponds to the opposite surface, at the bottom of the system.}
\label{Ech32Graph}

\end{figure}

The effect of temperature has been first investigated at 300K. Plasticity occurs for a 6.6\% elongation, showing that thermal activation significantly reduces the elastic limit. Another important difference due to temperature is the geometry of the formed defect. Instead of a straight dislocation, a dislocation half-loop forms and propagates throughout the crystal (Fig.~\ref{Ech32Atom1}). As expected, the nucleation of a half-loop dislocation is thermally activated. Contrary to the 0K simulation, a dislocation has nucleated from only one step: no dislocation is emitted from the other surface step, which remains intact. Atomic displacements at different simulation times (Fig.~\ref{Ech32Graph}) indicates that this half-loop dislocation has a Burgers vector orthogonal to the step, the screw component being almost zero. The formed dislocation is then a Shockley partial, leaving a stacking fault in its path. Atomic displacements have been fitted with an arctan function, according to elasticity theory. This allows to monitor the position of the dislocation core defined as the maximum of the derivative, during the simulation.

Before the complete propagation of a dislocation, several half-loop embryos starting from both steps have been observed, appearing and disappearing (Fig.~\ref{Ech32Atom1}). Only one of them will eventually become large enough and propagate into the crystal (Fig.~\ref{Ech32Atom1}). This is related to the existence of a critical size for the dislocation formation, due to attractive interaction with the free surface. As the dislocation moves through the crystal and reaches the opposite surface, a trailing partial does not nucleate. Though it would significantly reduce the total energy of the system, especially in aluminum which have a high stacking-fault energy, this would require the crossing of a high energy barrier. On the contrary, the successive nucleation of dislocations in adjacent \{111\} can be achieved with a much lower energy barrier. So, although it relaxes less energy than a trailing partial would, this mechanism is more likely to be activated. This is what we obtained in most simulations, similar to the twinning mechanism proposed by Pirouz \cite{Pirouz1995}. The remaining smaller step on the top surface, as well as the step created by the emergent dislocation on the bottom surface, become privileged sites for the nucleation of other dislocations in adjacent \{111\} planes, leading to the formation of a twin. While sufficient stress remains, successive faulted half-loops will be formed in adjacent planes, increasing the thickness of the twin. After 76~ps, the crystal structure does not evolve anymore. The plastic deformation is then characterized by a micro-twin (Fig.~\ref{Ech32Atom1}), located around the previous position of the step, with an extension of eight atomic planes, and delimited by two twin boundaries whose total energy equals the energy of an intrinsic stacking fault.

\begin{figure}

  \centering
  \includegraphics[width=0.5\linewidth]{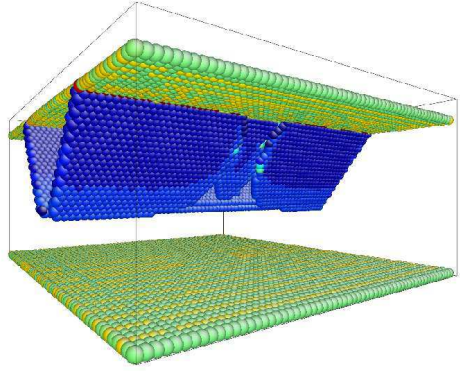}
  \caption{Dislocations nucleated in a crystal with one straight step, and one irregular, elongated by 10\% at 0K.}
\label{Ech05Atom1}

\end{figure}

We have also investigated how the dislocation formation process is modified in the case of irregular steps. We used a crystal with the same geometry, except that 10 consecutive atoms have been removed from one surface step edge (see Fig.~\ref{CrystalGeom}), creating two step kinks between which lies a notch. The other step remains straight. First, at 0K, no defect is obtained up to 10\% elongation, beyond which plasticity occurs. This elastic limit is similar to the one obtained for the system with perfect steps. Moreover, nucleated dislocations are also Shockley partials with a Burgers vector orthogonal to the step, and are emitted from both surface steps, despite the system asymmetry. However, two dislocations have been formed from the irregular step. In fact, a second partial nucleates and propagates in the \{111\} plane passing through the notch (Fig.~\ref{Ech05Atom1}). Both dislocations remain in their respective glide plane, leaving two stacking faults. This suggests that kinks are strong anchors for dislocations. Nevertheless, at 0K, it seems they have a negligible effect on the elastic limit, or regarding the nature of the nucleation event.

\begin{figure}

  \centering
  \includegraphics[width=\linewidth]{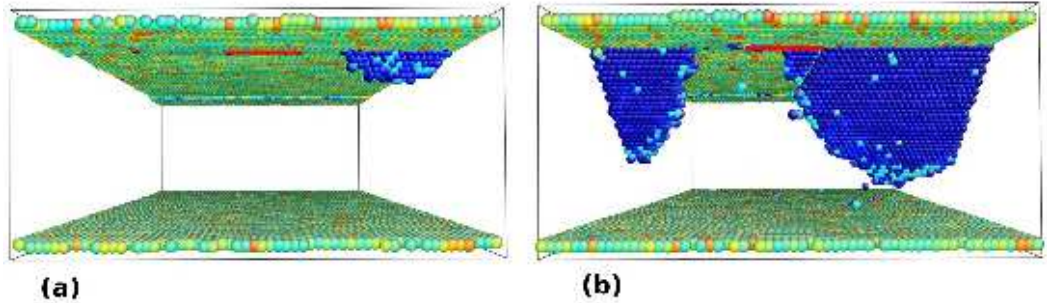}
  \caption{Evolution of the aluminum crystal with an irregular step, under a 6.6\% elongation at 300K. Same color and time conventions as in Fig.~\ref{Ech32Atom1}. The position of the notch is highlited in red. (a) After 7.4~ps, a faulted half-loop dislocation nucleates in the original \{111\} plane. (b) At 10~ps, another dislocation is emitted in the adjacent \{111\} plane, passing through the notch.}
\label{Ech35Atom1}

\end{figure}

At 300K and for the same geometry, the elastic limit is reached for a 6.6\% elongation, i.e. similar to the crystal with straight steps. Again, it suggests that irregular steps have no effect on the elastic limit. The dislocation half-loop does not nucleate from a step kink, but about 15 atomic planes away from it (Fig.~\ref{Ech35Atom1}a). It propagates into the crystal, but stays anchored to the kink, which acts like an obstacle to the movement. Then, another dislocation nucleates in the adjacent \{111\} plane, within the notch (Fig.~\ref{Ech35Atom1}b). Another simulation on a similar system leads to a dislocation nucleation from the straight step, despite the presence of a kinked step. These results show that kinks are not preferential sites for nucleation. It can be explained because step kinks are 0-D defects, contrary to straight steps, what prevents to initiate 1-D defects such as dislocations. After the first nucleation, the twinning mechanism, already described above, is observed. At about 70~ps, the formed twin cannot be distinguished from the one obtained in the crystal with straight steps. Finally, there is no indication left whether the step was initially irregular or not.

Molecular dynamics simulations have been used to investigate the influence of temperature and of step geometry on the first stages of plasticity in f.c.c. aluminum slabs. Surface steps were shown to be privileged sites for the nucleation of dislocations, significantly reducing the elastic limit compared to a perfect surface. Simulations with straight surface steps have revealed that only straight $90^{\circ}$ dislocations could nucleate at 0K. Temperature reduces the elastic limit, and makes possible the nucleation of faulted dislocation half-loops. Due to the system geometry and the strain orientation, only Shockley partials were obtained. Successive nucleation of partials in adjacent \{111\} planes are observed, similar to the twinning mechanism described by Pirouz in semiconductors. Simulations with an irregular step have shown that a kink is not a systematic site for nucleation. Instead, half-loops have been obtained from a straight portion of the step. The kinks introduced along a step seem to be strong anchor points for dislocations, making their motion more difficult along the step.

During all simulations including temperature, several dislocation half-loop embryos were observed before one eventually becomes large enough and propagates into the crystal. Calculations are in progress to determine the critical size a half-loop must reach to fully propagate. To determine the activation energy of the nucleation from surface steps, two methods may be used. First, the nudged elastic band method \cite{Jonsson1998,Henkelman2000,Henkelman2000a}, applied to the nucleation and propagation of a half-loop, would provide the minimum energy path for this event. Second, by performing several simulations at a given strain, one would obtain the average nucleation time as a function of temperature, thus allowing determination of the activation energy from Arrhenius plots. The dislocations speeds, as well as the size and shape of the dislocation half-loops, can be expected to depend on temperature, which will also be investigated through simulations. As a sequel to the nucleation event, several scenarii were observed. The twinning mechanism is supposed to be in competition with the nucleation of a trailing partial, which requires the crossing of a higher energy barrier. However this last mechanism was obtained during a simulation, showing it is still possible. More investigations would allow to determine the exact dependancy on temperature, strain, or other parameters.


P. Hirel's PhD work is supported by the R\'egion Poitou-Charentes. We greatly aknowledge the Agence Nationale de la Recherche for financing the project (number ANR-06-blan-0250).



\end{document}